\documentclass[preprint, pre, aps, 11pt, showpacs, superscriptaddress, showkeys]{revtex4-1}

\usepackage{amsmath,amssymb}
\usepackage{graphicx}
\usepackage[top=2in, bottom=1.5in, left=1in, right=1in]{geometry}
\usepackage{array}
\usepackage[mathscr]{euscript}
\usepackage{color}

\newcommand{\iu}{\mathrm{i}} 
\newcommand{\dd}{\mathrm{d}} 
\newcommand{\ee}{\mathrm{e}} 
\newcommand{\Kast}{K^\ast}
\newcommand{\Tc}{T_\mathrm{c}}
\newcommand{\Kc}{K_\mathrm{c}}
\newcommand{\Kcast}{K_{\mathrm{c}}^\ast}
\newcommand{\kB}{k_\mathrm{B}}

\newcommand{\fint}{f_\text{int}}
\newcommand{\meW}{\mathfrak{W}\hspace{0.3mm}}
\newcommand{\mew}[3]{\mathfrak{W}\hspace{0.3mm}_{#1,#2}^{\left(#3\right)}}
\newcommand{\ctgh}{\coth}
\newcommand{\sign}{\operatorname{sign}}
\newcommand{\rmi}{\mathrm{i}}
\newcommand{\partf}{\mathcal{Q}}
\newcommand{\fe}{\mathcal{F}}
\newcommand{\Tr}{\operatorname{Tr}}
\newcommand{\mS}{\mathbf{S}}
\newcommand{\xib}{\xi_\text{b}}

\newcommand{\opV}{\mathscr{V}}

\newcommand{\opI}{\mathscr{I}}
\newcommand{\opPM}{\mathscr{P}_M}
\newcommand{\opP}{\mathscr{P}}
\newcommand{\opR}{\mathscr{R}}
\newcommand{\opC}{\mathscr{C}}
\newcommand{\opA}{\mathscr{A}}
\newcommand{\opB}{\mathscr{B}}
\newcommand{\opD}{\mathscr{D}}
\newcommand{\opG}{\mathscr{G}}
\newcommand{\BesselK}{\mathcal{K}}
\newcommand{\shift}[2]{{#2}^{\left(#1\right)}}
\newcommand{\funcA}{\mathcal{A}}
\newcommand{\funcB}{F_{\text{sd}}}

\begin{document}
\date{\today}
\title{Critical Casimir forces between defects in the 2D Ising model}
\author{P. Nowakowski}
\affiliation{Max--Planck--Institut f\"ur Intelligente Systeme, Heisenbergstr.\ 3, D-70569 Stuttgart, Germany}
\affiliation{IV.\ Institut f\"ur Theoretische Physik, Universit\"at Stuttgart, Pfaffenwaldring 57, D-70569 Stuttgart, Germany}
\author{A. Macio\l{}ek}
\affiliation{Institute of Physical Chemistry, Polish Academy of Sciences, Kasprzaka 44/52, PL-01-224 Warsaw, Poland}
\author{S. Dietrich}
\affiliation{Max--Planck--Institut f\"ur Intelligente Systeme, Heisenbergstr.\ 3, D-70569 Stuttgart, Germany}
\affiliation{IV.\ Institut f\"ur Theoretische Physik, Universit\"at Stuttgart, Pfaffenwaldring 57, D-70569 Stuttgart, Germany}
\begin{abstract}
An exact statistical mechanical derivation is given of the critical Casimir interactions between two defects in a planar lattice--gas Ising model. Each defect is a group of nearest--neighbor spins with modified coupling constants. Such a system can be regarded as a model of a binary liquid mixture with the molecules confined to a membrane and the defects mimicking protein inclusions embedded into the membrane. As suggested by recent experiments, certain cellular membranes appear to be tuned to the proximity of a critical demixing point belonging to the two--dimensional Ising universality class. Therefore one can expect the emergence of critical Casimir forces between membrane inclusions. These forces are governed by universal scaling functions, which we derive for simple defects. We prove that the scaling law appearing at criticality is the same for all types of defects considered here. 
\end{abstract}
\maketitle

\newlength{\figwidth}
\setlength{\figwidth}{0.48\textwidth}

\section{Introduction}
\label{sec:int}

The idea that thermal fluctuations of a medium can lead to effective interactions between immersed objects was first pointed out by Fisher and de Gennes~\cite{Fisher78}. Since the origin of these forces is very similar to the Casimir force induced by the quantum fluctuations of electromagnetic fields~\cite{Casimir48}, they are called critical Casimir forces (CCF). 

When the system is close to its critical point, the thermal fluctuations of the corresponding order parameter are strong and long--ranged so that CCF can compete with and even dominate non--critical background forces. Due to universality of critical phenomena in confinement, these forces do not depend on microscopic details of the system and can be described in terms of universal scaling functions~\cite{Krech94, *Krech99, Brankov00}. They depend only on dimensionless scaling variables, which are typically ratios involving the geometric parameters characterizing the objects immersed in the fluctuating medium, the separation between them, and the bulk correlation length, which diverges upon approaching the critical point. These scaling functions describe all systems which belong to the same bulk and surface universality class. For over three decades CCF and their scaling functions have been studied experimentally, theoretically, and numerically for various bulk and surface universality classes~\cite{[{For a recent review see }]Gambassi09}. The CCF have been measured for wetting films of $^4\mathrm{He}$ and $^3\mathrm{He}$--$^4\mathrm{He}$ mixtures near their $\lambda$--transition and their tricritical point of the bulk system, respectively~\cite{Garcia99, Garcia02, Ueno03, Ganshin06}. Also experiments with binary liquid mixtures near the mixing--demixing transition have been performed for wetting films~\cite{Fukuto05, Rafai07} and for the sphere--plate geometry~\cite{Hertlein08, Soyka08, Gambassi09a, Nellen09, Troendle11}. These experimental findings agree quantitatively with corresponding theoretical analyses~\cite{Krech92, *Krech92a, Krech97, Brojan98, *Brojan08, Maciolek06} and Monte Carlo simulations~\cite{Vasilyev07, *Vasilyev09, Hucht07, Hasenbusch10, Hasenbusch13}.

Recently, a phenomenon, which has been interpreted as the occurrence of a critical point, was observed in giant plasma membrane vesicles isolated from living cells~\cite{Veatch08}. This phase transition is similar to a critical demixing of a two--dimensional (2D) binary liquid mixture and belongs to the 2D Ising universality class. The same phenomenon occurs in certain lipid membranes~\cite{Honerkamp-Smith08, *Heinrich08}. These are 2D liquids consisting of two (or more) components such as cholesterol, saturated, and unsaturated lipids, which undergo separation into two liquid phases, one rich in the first two components and the other rich in the third~\cite{Veatch05}. Lipid membranes serve as model systems for cell plasma membranes~\cite{Lingwood10}. The occurrence of a continuous demixing transition in membranes would definitely give rise to fluctuation--induced, effective forces acting between proteins embedded in membranes as inclusions. Thus, studying these forces contributes to the understanding of how these biological cells work.

Another type of Casimir--like interactions between membrane inclusions appears due to the restriction of the thermal fluctuations of the local shape of a membrane caused by the presence of these inclusions~\cite{Goulian93, Kardar99, Bitbol10, Yolcu11, Lin11, Semrau09, Abraham07}. (This resembles capillary--wave--induced effective interactions among colloids floating at fluid interfaces~\cite{Lehle06}.) Such membrane--fluctuation--induced interactions are attractive and their form depends sensitively on the shapes of the inclusions and the membrane rigidity. 

In the present study we model the inclusions as defects of a finite size on the infinite square lattice and calculate CCF analytically by using exact diagonalization of the transfer matrix for the 2D Ising model~\cite{Onsager44, Kaufman49}. This allows us to determine an exact expression for the interaction free energy between two finite--sized defects of arbitrary shape (in the absence of the bulk ordering field). In practice, tractable expressions can be obtained for defects small in size and of simple shapes. For such defects, these expressions can be further analyzed in order to determine their asymptotic behavior for large separations $\ell$ between the defects at \textit{any} fixed temperature $T$ and thus to calculate the scaling function of the CCF. In fact, in the scaling limit the functional form of the CCF can be determined for defects of arbitrary shape. It turns out that it has a very simple product structure, in which the terms depending only on the properties of defects factor out from the universal scaling part.

The critical Casimir interaction between two inclusions in the 2D Ising square lattice has been studied also by other approaches. In Ref.~\cite{Machta12} conformal field theory (CFT) and Monte Carlo (MC) simulations have been applied for two disclike objects. Contrary to the present approach, CFT is limited to the bulk critical point. In principle, in the limit $\ell \to \infty$, i.e., in which the details of the inclusions do not matter, the results of the present microscopic model at the critical point $T=T_c$ should coincide with those from the coarse--grained CFT approach. In the limit of large $\ell$, the CFT expression for the interaction free energy between two discs has a similar product form as the one we obtain here. Our calculations show that the same structure persists off critical point. The MC simulation data presented in Ref.~\cite{Machta12} are also very limited. Apart from the results at the bulk critical point, the interaction free energy between two inclusions as a function of their separation was determined for only three other temperatures, all above $\Tc$. No scaling function has been determined because the system sizes for which the simulations have been performed are too small for reaching the scaling limit. In Ref.~\cite{Zubaszewska13}, also the off--critical behavior of CCF has been studied by using very accurate (albeit not exact) numerical density--matrix renormalization--group techniques. However, the scaling functions of CCF between two discs have been determined within the Derjaguin approximation~\cite{Derjaguin34}, which is valid only if the distance between the disclike objects is much smaller than their radius. This is the limit opposite to the one considered here.

Generally, if two parallel $\left(d-1\right)$--dimensional plates are immersed in a $d$--dimensional fluid at a distance $\ell$, at the bulk critical point $\Tc$ the interaction free energy decays as $\ell^{-\left(d-1\right)}$, which follows from finite--size scaling and dimensional analysis~\cite{Fisher78}. For 2D plates immersed in a three--dimensional fluid ($d=3$) this leads to the well known decay $\sim \ell^{-2}$ of the interaction free energy and the decay $\sim \ell^{-3}$ of the CCF.

In the case of two strictly finite--sized defects the argument presented in Ref.~\cite{Fisher78} does not allow one to determine the dependence on $\ell$ of the interaction free energy at the critical point. Instead, from liquid state theory \cite{Hansen76} it is known that the effective pair potential $U(\left|\mathbf{r}_1-\mathbf{r}_2\right|=r)$ for two solute particles 1 and 2 at $\mathbf{r}_1$ and $\mathbf{r}_2$, respectively, in a bulk solvent is given by $U(r) = -k_BT \ln g(r)$ in the limit of dilute suspensions, where $g(r)$ is the pair distribution function. In spin lattice systems, the presence of a solute particle corresponds to having a cluster of fixed spins. Accordingly, the effective pair interaction potential is governed by two--point correlation functions of the Ising model. This correspondence has been confirmed in Ref.~\cite{Burkhardt95} by explicit calculations for two small spheres immersed in a critical fluid. In these calculations, the small--sphere expansion (which amounts to a kind of short--distance expansion) has been used to express the interaction free energy in terms of a correlation function of a local operator such as the order parameter or the energy--density of a solvent, the latter depending on the surface universality classes of both spheres. The coefficient of this expansion is related to the amplitudes of the leading decay of the bulk two--point correlation function and of the profile in a half--space of the corresponding local operator. For symmetry--breaking boundary conditions on both spheres the interaction free energy decays as $\ell^{-2\beta/\nu}$, like the bulk spin---spin correlation function of the Ising model, whereas for symmetry--preserving boundary conditions on both spheres the interaction free energy decays as $\ell^{-2\left(d-1/\nu\right)}$, like the bulk energy--density---energy--density correlation function of the Ising model. Here $\ell$ is the distance between the centers of the spheres, $d$ is the bulk dimension of the system, and $\beta$ and $\nu$ are standard bulk critical exponents. We expect that even for defects of anisotropic shapes, such as needle-- or L--shaped objects, the leading behavior of the interaction free energy at large distances is governed by the appropriate correlation function. A needle embedded in a bounded 2D Ising strip at bulk criticality was studied in Ref.~\cite{Vasilyev13} by MC simulations and CFT as an instructive paradigm for investigating the universal orientation--dependent interactions between nonspherical colloidal particles immersed in a critical solvent.

In this study we consider the 2D Ising model ($d=2$) on a square lattice with two defects. Inside each defect the couplings between spins are modified. Because this type of defect preserves the symmetry with respect to changing the sign of all spins, the defect amounts to a symmetry--preserving boundary conditions. For the 2D Ising model $\nu=1$~\cite{Pelissetto02}. Thus if the above result for spheres holds also for non--spherical defects in the small particle limit, the interaction free energy is expected to decay as $\ell^{-2}$ for large distances. 

The paper is organized as follows. In Sec.~\ref{sec:mod} we define the model and describe the type of defects considered here. In Sec.~\ref{sec:ccf} we introduce the CCF between these defects and provide a formula which we use to calculate it. In Sec.~\ref{sec:sd} we study the interaction free energy and the force for the simplest possible defects. We derive their thermodynamic properties in the scaling limit. In Sec.~\ref{sec:cd} we present our results for arbitrary shapes of the defects. In Sec.~\ref{sec:circ} we compare our results with those obtained from CFT and reported in the literature. Finally, we summarize our results in Sec.~\ref{sec:sc}. Technical details are presented in the Appendices \ref{app:A}--\ref{app:E}. We recall the original solution of the Ising model (Appendix~\ref{app:A}), describe our method to introduce defects on the lattice (Appendix~\ref{app:B}), prove formulae needed to calculate the CCF (Appendix~\ref{app:C}), and describe our approach to calculate it (Appendix~\ref{app:D}). In the last Appendix \ref{app:E} we prove the general formula for the CCF in the scaling limit.

\section{Model}
\label{sec:mod}

\begin{figure}[t]
\includegraphics[width=\figwidth]{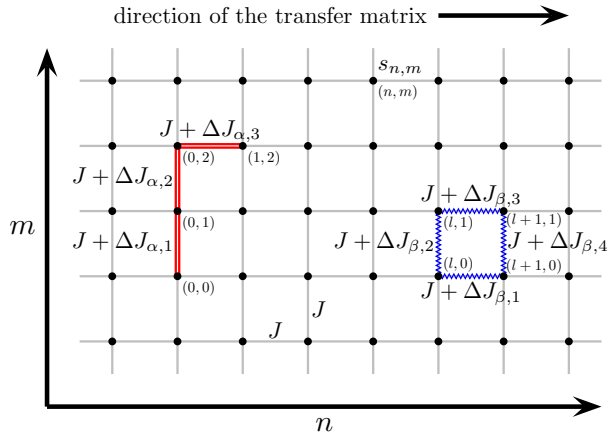}	
\caption{\label{fig:Sys1} Lattice model with two defects. Each spin $s_{n,m}$ of the 2D Ising model is interacting with nearest neighbors via a coupling constant $J$. These bonds are denoted by gray lines. Modified couplings are indicated by red double lines and blue zigzag lines for the first ($\alpha$) and the second ($\beta$) defect, respectively. The two defects are separated by a distance $l$ as shown in the figure. Within our model, the shape of each defect is arbitrary; accordingly, the figure presents one possible case.} 
\end{figure}

We consider the 2D Ising model on a square lattice of $M$ rows and $N$ columns. A state of the spin located in the $m$-th row and the $n$-th column is denoted by $s_{n,m}=\pm 1$, where $n=0,1,2,\ldots, N-1$ and $m=0,1,2,\ldots,M-1$. The lattice constant is set to $a=1$ and we apply periodic boundary conditions in both directions: $s_{N,m}\equiv s_{0,m}$ and $s_{n,M}\equiv s_{n,0}$. Neighboring spins interact with a ferromagnetic coupling constant $J>0$. In order to study the CCF we introduce two defects. Accordingly, the Hamiltonian of the system is 
\begin{equation}
\mathcal{H}=\mathcal{H}_\text{Ising}+\mathcal{H}_\text{defect 1}+\mathcal{H}_\text{defect 2},\\
\end{equation}
where $\mathcal{H}_\text{Ising}$ is the standard Hamiltonian of the Ising model
\begin{equation}\label{HIsing}
\mathcal{H}_\text{Ising}=-J\sum_{n=0}^{N-1}\sum_{m=0}^{M-1}\left(s_{n,m}s_{n+1,m}+s_{n,m}s_{n,m+1}\right),
\end{equation}
and $\mathcal{H}_\text{defect $i$}$ describes the modification of the standard Hamiltonian due to the $i$-th defect. In the present context a defect means modified coupling constants between several pairs of neighboring spins. The values of these modifications are denoted by $\Delta J_{i,j}$, where $i=\alpha,\beta$ marks the first and the second defect, respectively (see Fig.~\ref{fig:Sys1}), and $j$ labels pairs of nearest--neighbor spins with a modified bond in the defect. We shall also use the vector notation
\begin{equation}
\mathbf{\Delta J}_i=\left(\Delta J_{i,1},\Delta J_{i,2}, \Delta J_{i,3}, \ldots\right).
\end{equation}
We note that the total coupling between two spins belonging to one of the defects bond is $J+\Delta J$. We consider all possible values of $\Delta J$, including $\pm \infty$, in which case the neighboring spins are forced to be in the same or the opposite state. In the present study we consider only two defects of certain, possibly different, shapes. The size of each defect can be arbitrary but the number of modified couplings must be finite. The position of the first defect is taken to be fixed, such that one of its sites is at the origin $\left(0,0\right)$. The second defect encompasses the lattice sites belonging to the columns with the index $l+n$, with $n$ taking several integer values, so that the distance between defects is $\ell=a l$ (where $a$ is the lattice constant). Changing $l$ shifts the second defect in the horizontal direction without changing its shape. If the defects span over several columns of the system, the distance $l$ between the defects depends on an arbitrary choice of the position of a reference point. However, for large $l$ the leading order asymptotic behavior of the interaction free energy of defects is independent of this ambiguity. An example of such defects is shown in Fig.~\ref{fig:Sys1}.

All results reported here are calculated in the thermodynamic limit. For simplicity we use the limit $N\to\infty$ followed by $M\to\infty$ instead of the standard way $M,N\to\infty$ with $N/M$ fixed. The results obtained for the critical Casimir interaction are the same for both versions of the thermodynamic limit (see Appendix~\ref{app:C}).

\section{Critical Casimir force}\label{sec:ccf}

The free energy of the system with two (2) defects is given by
\begin{equation}
\fe_{2}\left(T,l,\mathbf{\Delta J}_\alpha, \mathbf{\Delta J}_\beta,M,N\right)=-\kB T \ln\partf\left(T,l,\mathbf{\Delta J}_\alpha, \mathbf{\Delta J}_\beta,M,N\right),
\end{equation}
where $\partf=\sum_{\left\{s_{n,m}\right\} \exp\left[-\mathcal{H}/\left(\kB T\right)\right]}$ is the partition function. This free energy decomposes as follows:
\begin{multline}
\fe_{2}\left(T,l,\mathbf{\Delta J}_\alpha, \mathbf{\Delta J}_\beta,M,N\right)=N M
f_{\mathrm{b}}\left(T\right)+f_\alpha\left(T,\mathbf{\Delta
    J}_\alpha\right)+f_\beta\left(T,\mathbf{\Delta J}_\beta\right)\\
+\fint\left(T,l,\mathbf{\Delta J}_\alpha, \mathbf{\Delta
    J}_\beta\right)+f_{\text{finite},2}\left(T,l,\mathbf{\Delta J}_\alpha, \mathbf{\Delta
    J}_\beta,M,N\right)+f_0\left(T\right),
\end{multline}
where $f_{\mathrm{b}}$ is the bulk free energy density, $f_{\alpha}$ ($f_\beta$) is the point contribution to the free energy stemming from the individual defect $\alpha$ ($\beta$), $\fint$ is the free energy of interaction between the two defects mediated by the unbounded system, $f_{\text{finite},2}$ is the finite size correction to the free energy, and $f_0\left(T\right)$ is a $M$-- and $N$--independent contribution which depends on how the thermodynamic limit is taken (see Appendix~\ref{app:C}). By construction $f_{\text{finite},2}$ vanishes in the limit $N,M\to\infty$ and $\fint$ decays to zero for $l\to\infty$.

The CCF are defined as the negative gradient of the free energy of interaction between objects immersed in the system. Since here the distance between the defects can take only integer values, we use a difference instead of a derivative:
\begin{equation}\label{defFCas}
F_{\text{Cas}}\left(T,l+\frac{1}{2},\mathbf{\Delta J}_\alpha, \mathbf{\Delta J}_\beta\right)=-\left[f_\text{int}\left(T,l+1,\mathbf{\Delta J}_\alpha,\mathbf{\Delta J}_\beta\right)-f_\text{int}\left(T,l,\mathbf{\Delta J}_\alpha,\mathbf{\Delta J}_\beta\right)\right].
\end{equation}

In order to determine the free energy we calculate the partition function $\mathcal{Q}$ of the system using the transfer matrix method. We use the transfer operator of the 2D Ising model on a square lattice with no defects. The construction and the diagonalization of this operator is recalled in Appendix~\ref{app:A}. The two defects are taken into account by inserting between the transfer operators special operators $\opD_\alpha\left(T,\mathbf{\Delta J}_\alpha\right)$ and $\opD_\beta\left(T,l,\mathbf{\Delta J}_\beta\right)$ which modify the couplings between the spins belonging to a defect. The method of constructing these operators is described in Appendix~\ref{app:B}.

In Appendix~\ref{app:C} we derive the following expression for the interaction free energy (with the limit $N\to\infty$ already carried out):
\begin{multline}\label{fint}
\fint\left(T,l,\mathbf{\Delta J}_\alpha, \mathbf{\Delta J}_\beta\right)/\left(\kB
  T\right)=\\-\ln\frac{\lim_{M\to\infty}\left<0\middle| \opD_\alpha\left(T,\mathbf{\Delta J}_\alpha\right)
    \opD_\beta\left(T,l,\mathbf{\Delta J}_\beta\right)\middle|
    0\right>}{\lim_{M\to\infty}\left<0\middle|\opD_\alpha\left(T,\mathbf{\Delta
          J}_\alpha\right)\middle| 0\right>\lim_{M\to\infty}\left<0 \middle
      |\opD_\beta\left(T,l,\mathbf{\Delta J}_\beta\right)\middle| 0\right>},
\end{multline}
where $\left|0\right>$ is the eigenvector of the transfer matrix corresponding to the highest eigenvalue (see \eqref{eq:A:vacuum}). Both the eigenvector and the operators $\opD_i$ depend on $M$. In Appendix~\ref{app:D} we derive some formulae which are useful for calculating the matrix elements in the above equation.

\section{Simple defects}\label{sec:sd}

\begin{figure}[t]
\includegraphics[width=\figwidth]{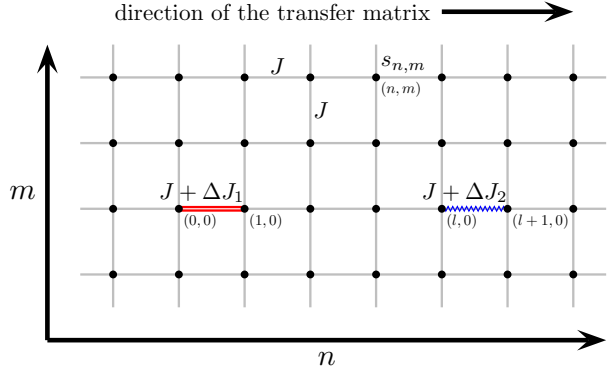}
\caption{\label{fig:Sys2} Two simple defects. Each of them is a modified coupling between two spins in the same row. The two defects are separated by a distance $l$.}
\end{figure}

We start our analysis from the simplest case, in which each defect consists of only a single modified coupling between spins in one row, and both defects are in the same row. This system is shown in Fig.~\ref{fig:Sys2}. The contributions to the Hamiltonian, which describe the defects, are 
\begin{equation}\label{simple_def_Hamiltonian}
\mathcal{H}_\text{defect 1}=-\Delta J_1 s_{0,0}s_{1,0},\qquad
\mathcal{H}_\text{defect 2}=-\Delta J_2 s_{l,0}s_{l+1,0}, \qquad l\geqslant 1,
\end{equation}
where $\Delta J_1$ and $\Delta J_2$ are the modifications of the coupling constant between the two spins belonging to each defect. Following the method described above we assign an operator to each defect (see Appendix~\ref{app:B}):
\begin{equation}
\opD_\alpha\left(T,\Delta J_1\right)=\opR_0\left(\Delta J_1\right), \qquad
\opD_\beta\left(T,l,\Delta J_2\right)=\shift{l}{\opR}_0\left(\Delta J_2\right),
\end{equation}
where the operators $\opR_0$ and $\shift{l}{\opR}_0$ are given by Eqs.~\eqref{defR} and \eqref{eq:B:ndef}. Using Eq.~\eqref{fint} and after some algebra, we obtain the following formula for the free energy of interaction:
\begin{multline}\label{fintsimple}
\fint\left(T,l,\Delta J_1,\Delta J_2\right)/\left(\kB T\right)=-\ln\Big\{1-\sinh\Delta K_1
  \sinh\Delta K_2
  \left[\left(\mew{0}{0}{l}\right)^2+\left(\mew{0}{1}{l}\right)^2
  \right]\\
\times\left[\sinh\left(2K+\Delta K_1\right)-\mew{0}{1}{0} \sinh\Delta
    K_1\right]^{-1}\left[\sinh\left(2K+\Delta K_2\right)-\mew{0}{1}{0} \sinh\Delta
    K_2\right]^{-1}\Big\}
\end{multline}
where $\Delta K_i=\Delta J_i/\left(\kB T\right)$ and the matrix elements $\mew{j}{k}{l}$ are given by Eq.~\eqref{meW}. From this result the CCF can by calculated via Eq.~\eqref{defFCas}. For all non--zero temperatures, the force is negative (i.e., attractive) if $\Delta J_1$ and $\Delta J_2$ have the same sign, and positive (i.e., repelling) if $\Delta J_1\Delta J_2<0$. Within the present model with short--ranged interactions, there is no force between defects at zero temperature or, trivially, if one of the defects is removed (i.e., $\Delta J_1=0$ or $\Delta J_2=0$). 

\begin{figure*}
\includegraphics[width=\figwidth]{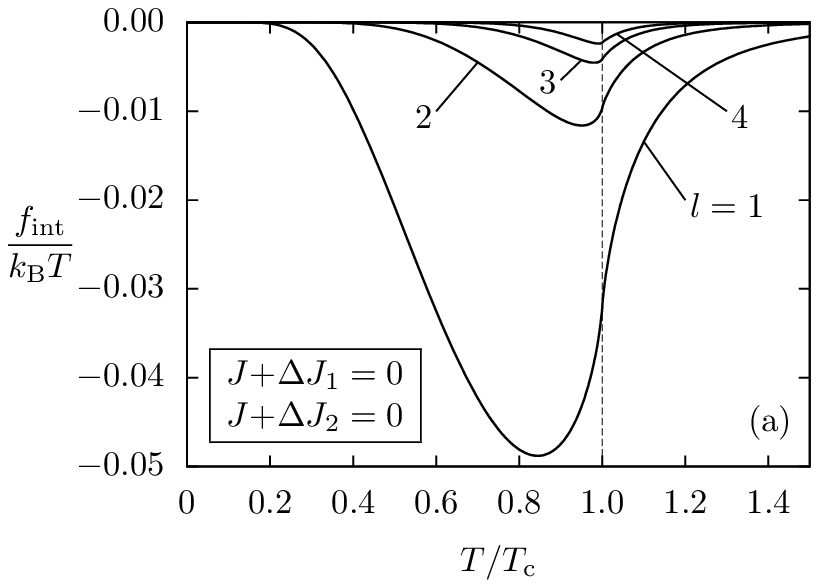} \hfill \includegraphics[width=\figwidth]{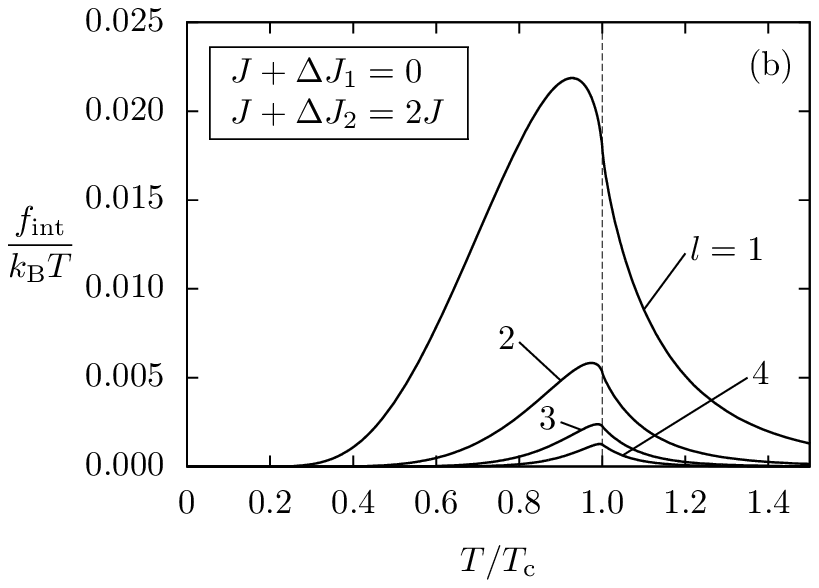}
\caption{\label{fig3} Interaction free energy $\fint$ for two simple defects as a function of temperature for four separations $l$ between them. (a) $J+\Delta J_1=J+\Delta J_2=0$, (b) $J+\Delta J_1=0$, $J+\Delta J_2=2J$.}
\end{figure*}

Figure~\ref{fig3} shows the interaction free energy $\fint$ (see Eq.~\eqref{fintsimple}) as a function of temperature for four separations $l$. Near $\Tc$, $\fint$ exhibits a non--analyticity $\sim\mathrm{const}+t\ln\left| t\right|$, where $t=\left(T-\Tc\right)/\Tc$. The position of the extremum of $\fint$ is always located below $\Tc$ and approaches the bulk critical temperature $\Tc$ upon increasing $l$.

The behavior of $\fint$ for fixed $l$ and $t\to 0$ agrees with the result obtained in Ref.~\cite{Wu15} that close to the critical point the excess internal energy of a single defect behaves like the specific heat of a homogeneous system. In the case of the 2D Ising model, where the specific heat diverges $\sim \ln\left|t\right|$, this leads to the non--analyticity $\sim t\ln\left|t\right|$ of the excess free energy of a single defect. Finally, because in the limit $t\to 0$ two defects fixed at a distance $l$ can be considered as one big defect, $\fint$ is expected to exhibit the same type of non--analyticity (see, cf, the discussion after Eq.~\eqref{funcG}). This is in line with the observation reported above.

In order to study the dependence of $\fint$ on the distance $l$ we focus on the leading order term in Eq.~\eqref{fintsimple} in the limit $l\to\infty$ at fixed temperature $T\neq\Tc$:
\begin{multline}\label{fint_exp}
\fint\left( T,l,\Delta J_1,\Delta J_2\right)/\left(\kB
  T\right)=\\
-\funcA\left(\Delta J_1,T\right)\funcA\left(\Delta J_2,T\right)\frac{\exp\left(-4 l
    \left|K-\Kast\right|\right)}{2 \pi l^2}+\mathrm{O}\left[\frac{\exp\left(-4 l
    \left|K-\Kast\right|\right)}{l^3}\right],
\end{multline}
where we have introduced
\begin{equation}\label{F1}
\funcA\left(\Delta J_i,T\right)=\sinh \Delta K_i\sinh 2K\Big/\left[\sinh\left(2K+\Delta
    K_i\right)-\mew{0}{1}{0}\sinh\Delta K_i\right], \quad i=1,2.
\end{equation}
Concerning the dual coupling $\Kast$ see Appendix~\ref{app:A} (Eq.\ \eqref{Kast}). The interaction free energy factorizes into three terms: the first factor depends on temperature $T$ and on the strength $\Delta J_1$ of the coupling in the first defect, the second factor depends on $T$ and the coupling $\Delta J_2$ in the second defect, and the third factor depends on $T$ and the distance between defects. The lengthscale $\left|4K-4\Kast\right|^{-1}$ of the exponential decay of the interaction free energy for $T<\Tc$ is equal to the bulk correlation length $\xib\left(T\right)$, which controls the spatial exponential decay of the spin---spin correlation function; for $T>\Tc$ one has $\left|4K-4\Kast\right|^{-1}=\xib\left(T\right)/2$. Using Eq.~\eqref{defFCas} one finds that asymptotically in this limit $F_\mathrm{Cas}$ is proportional to $l^{-2}\exp\left(-4l\left|K-\Kast\right|\right)$. This exponential decay is characteristic for the CCF at large separations and off the critical point \cite{Krech94, Evans94}.

In order to study the CCF close to $\Tc$, we consider the scaling limit $l\to \infty$ and $T\to\Tc$ with the scaling variable
\begin{equation}\label{scalex}
x=t l/\xi_0^+\sim \sign\left(t\right)l/\xib(T)
\end{equation}
fixed, with $\xib(t\to 0^\pm) = \xi_0^\pm \left| t\right|^{-\nu}$, $\nu=1$, $\xi_0^+=\left[2 \ln\left(1+\sqrt{2}\right)\right]^{-1}$, and $\xi_0^-=\xi_0^+/2$ for the 2D Ising model and for the definition of $\xib$ given above. This is completely in line with previous studies of CCF in 2D Ising strips \cite{Evans94, Nowakowski08, Abraham10, Abraham13}. The results of our calculations show explicitly that the perturbation due to these defects is marginal, i.e., the corresponding scaling exponent is zero. This is not surprising because even an infinitely long defect line is a marginal perturbation of a 2D system \cite{Hanke00}.

In this limit the interaction free energy takes the form
\begin{equation}\label{scalesimple}
\fint\left(T,l,\Delta J_1,\Delta J_2\right)/\left(\kB
  \Tc\right)=-\funcB\left(\Delta J_1\right)
\funcB\left(\Delta J_2\right)
\frac{G\left(x\right)}{l^2}+\mathrm{O}\left(\frac{\ln l}{l^{3}}\right),
\end{equation}
where the index ``sd'' stands for ``simple defect'',
\begin{equation}\label{F2}
\funcB\left(\Delta J_i\right)=\left.\frac{\sqrt{2}\sinh \Delta K_i}{\sqrt{2}\cosh\Delta
    K_i+\sinh \Delta K_i}\right|_{T=\Tc},
\end{equation}
and
\begin{equation}\label{G}
G\left(x\right)=x^2 \left[\BesselK_1^2\left(\left|x\right|\right)-\BesselK_0^2\left(\left|x\right|\right)\right]/\pi^2.
\end{equation}
$\BesselK_0\left(x\right)$ and $\BesselK_1\left(x\right)$ are modified Bessel functions of the second kind. Also in the scaling limit, the interaction free energy factorizes into three terms. Two of them depend only on the modifications $\Delta J_1$ and $\Delta J_2$, respectively, of the bonds forming the two defects and one depends on the distance $l$ between the defects and on temperature via the scaling variable $x$. We note that, because $\mew{0}{1}{0}\left(T=\Tc\right)=1/\sqrt{2}$, one has $\funcB\left(\Delta J_i\right)=\funcA\left(\Delta J_i,T=\Tc\right)$ (see Eq.~\eqref{F1}).

As expected, close to $\Tc$ and to leading order in terms of $l^{-1}$ the interaction free energy depends on the microscopic details of each defect but its decay for large distances $l$ is the same as for two discs immersed in the 2D critical fluid \cite{Burkhardt95}. The function $G\left(x\right)$ has already been reported in the context of the 2D Ising model --- it is proportional to the energy--density---energy--density correlation function in the scaling limit \cite{Hecht67, Abraham94}. This confirms the relation between the energy--density---energy--density correlation function and the CCF between two defects of the type considered here. (Actually, it is possible to rederive Eq.~\eqref{scalesimple} using the results in Ref.~\cite{Hecht67}; this provides an independent check of our calculation.)

\begin{figure*}[t]
\includegraphics[width=\figwidth]{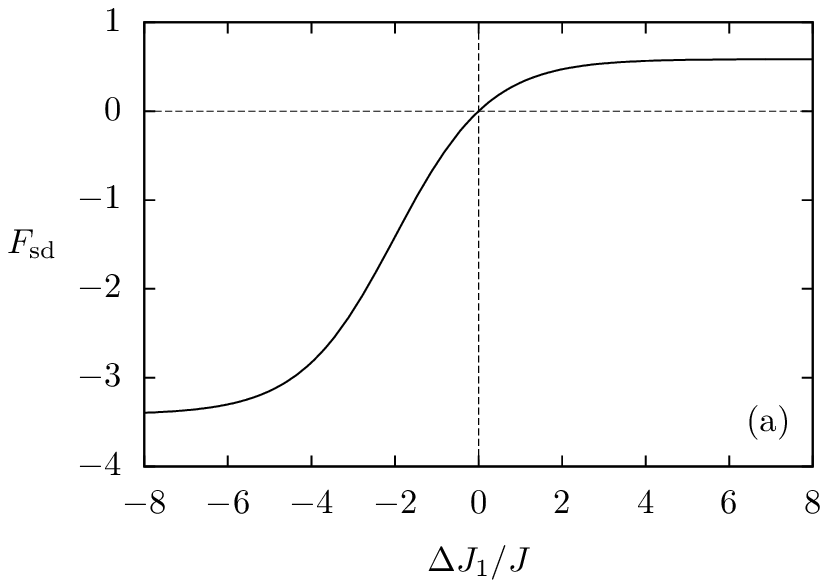} \hfill \includegraphics[width=\figwidth]{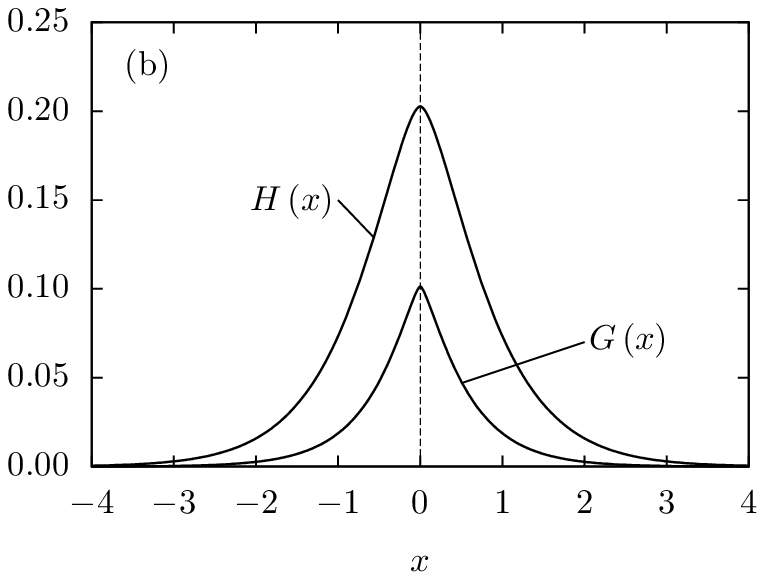}
\caption{\label{figF2GH}(a) The function $\funcB$ (Eq.~\eqref{F2}) which determines the strength of the CCF (Eq.~\eqref{FCas}). (b) Scaling functions $G\left(x\right)$ and $H\left(x\right)$ for the interaction free energy and the CCF, respectively.}
\end{figure*}

In the scaling limit the CCF $-\partial \fint/\partial l$ follows from Eq.~\eqref{scalesimple}:
\begin{equation}\label{FCas}
F_\text{Cas}\left(x,l,\Delta J_1,\Delta J_2\right)/\left(\kB
  \Tc\right)=-\funcB\left(\Delta J_1\right)\funcB\left(\Delta
  J_2\right)\frac{H\left(x\right)}{l^3}+\mathrm{O}\left(\frac{\ln l}{l^{4}}\right),
\end{equation}
where
\begin{equation}
H\left(x\right)=2G\left(x\right)-x \frac{\dd G\left(x\right)}{\dd
  x}=2x^2 \BesselK_1^2\left(x\right)/\pi^2.
\end{equation}
The functions $\funcB\left(\Delta J_i\right)$, $G\left(x\right)$, and $H\left(x\right)$ are presented in Fig.~\ref{figF2GH}. Both scaling functions $G\left(x\right)$ and $H\left(x\right)$ are symmetric around $x=0$. For large values of $\left| x \right|$ they decay exponentially and at $x=0$ they are continuous but non--analytic:
\begin{subequations}\label{funcG}
\begin{align}
\label{funcG1}G\left(x\to 0\right)&=1/\pi^2+\mathrm{O}\left(x^2\ln^2 \left|x\right|\right),\\
G\left(x\to\infty\right)&=\exp\left(-2x\right)\left[1/\left(2\pi\right)+\mathrm{O}\left(x^{-1}\right)\right],\\
H\left(x\to 0\right)&=2/\pi^2+\mathrm{O}\left(x^2\ln \left|x\right|\right),\\
H\left(x\to\infty\right)&=\exp\left(-2x\right)\left[ x/\pi+\mathrm{O}\left(1\right)\right].
\end{align}
\end{subequations}
Note that the non--analyticities of $G\left(x\right)$ at $x=0$ and of $\fint$ at $T=\Tc$ (see the paragraph preceding Eq.~\eqref{fint_exp}) are distinct. This can be traced back to different limiting procedures. The behavior of $\fint$ at $T=\Tc$ follows from the limit $t\to 0$ at a fixed distance $l$ between the defects. On the other hand, the scaling limit leading to Eqs.~\eqref{scalesimple} and~\eqref{funcG} amounts to the limits $t\to 0$ and $l\to\infty$ so that $l/\xib\left(T\right)$ remains constant. The fact, that these two limiting procedures yield different results, tells that in the scaling limit two defects cannot be considered as a single big defect.

\section{Richer lattice defects}\label{sec:cd}

In this section we consider lattice defects with richer structures. We allow each defect to have any finite number of modified coupling constants. In principle, for any particular shape of the defects it is possible to determine an exact expression for the interaction free energy using the same method as for the simple defects above. However, the complexity of the derivation increases rapidly with the size of defects. We have been able to calculate exact expressions only for systems with lattice defects containing up to three modified bonds.

In the scaling limit the functional form of the force can be studied without deriving explicit expressions. In Appendix~\ref{app:E} we prove that the interaction free energy between defects of arbitrary size, in the scaling limit $l\to\infty$ and $T\to\Tc$ with $x$ fixed, is given by 
\begin{equation}\label{sl}
\fint\left(T,l,\mathbf{\Delta J}_\alpha,\mathbf{\Delta J}_\beta \right)/\left(\kB
  \Tc\right)=-F_\alpha\left(\mathbf{\Delta
    J}_\alpha\right)F_\beta\left(\mathbf{\Delta J}_\beta\right)\frac{G\left(x\right)}{l^2}+\mathrm{O}\left(\frac{\ln l}{l^{3}}\right).
\end{equation}
In this expression, $F_\alpha\left(\mathbf{\Delta J}_\alpha\right)$ is a function which depends only on the structure and coupling constants in the defect $\alpha$, $F_\beta\left(\mathbf{\Delta J}_\beta\right)$ is the corresponding function reflecting the structure and the coupling constants in the second defect $\beta$, $x$ is the scaling variable introduced in Eq.~\eqref{scalex}, $G\left(x\right)$ is the scaling function given in Eq.~\eqref{G}, and $l$ is the distance between the defects measured in units of the lattice constant. Note that the functions $F_\alpha$ and $F_\beta$ are dimensionless and thus they can only depend on dimensionless ratios $\mathbf{\Delta J}_\alpha/\left(\kB \Tc\right)$ and $\mathbf{\Delta J}_\beta/\left(\kB \Tc\right)$, respectively.

The distance $l$ between defects depends on how the reference points in both defects are chosen; however, the leading term in the scaling law given by Eq.~\eqref{sl} does not depend on this choice. The expression is also valid if the two defects are separated in a direction which is not parallel to the underlying lattice directions. In this case the distance $l$ is taken to be the Euclidean distance between the reference points.

Calculating the expressions for $F_\alpha\left(\mathbf{\Delta J}_\alpha\right)$ is challenging, especially for large defects. For all shapes of defects which we have considered, the form of the factor $F_\alpha$ remains unchanged upon rotation by 90$^\circ$ or mirror reflection.

\section{Effective interaction between circular defects}\label{sec:circ}

\begin{figure}
\includegraphics[scale=1.0]{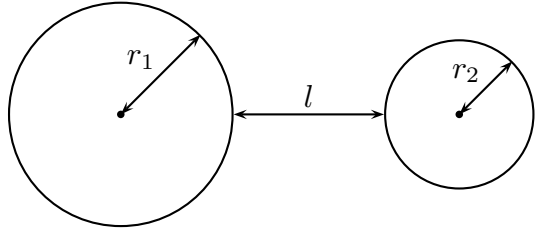}
\caption{\label{fig:CFTsys} Geometry of the systems considered in Ref.~\cite{Machta12}.}
\end{figure}

In Ref.~\cite{Machta12} a continuum version of the 2D Ising model has been considered. Using boundary CFT, the interaction free energy between circular inclusions of radii $r_1$ and $r_2$ separated by a distance $l$ was calculated at the critical temperature (see Fig.\ \ref{fig:CFTsys}). In this section we provide a comparison of these results with our corresponding expressions for the free energy in the scaling limit.

In Ref.~\cite{Machta12}, four distinct boundary conditions were considered but only one of them, called ``free--free'' (for which the discs do not couple to the order parameter) is not breaking the symmetry of reversing all spins simultaneously and thus can be compared with our results. To this end, first we expand the interaction free energy between two discs as obtained in Ref.~\cite{Machta12} in the scaling limit for large $l$ (see Eq.~(4) in Ref.~\cite{Machta12}).
\begin{equation}\label{eq:fintCFT}
f^{\mathrm{CFT}}_{\mathrm{int, circles}}\left(T=\Tc,l,r_1,r_2\right)/\left(\kB \Tc\right)=-\frac{r_1 r_2}{l^2}+\mathrm{O}\left(l^{-4}\right).
\end{equation}
On the other hand, from Eq.~\eqref{sl} for circular defects we expect at $T=\Tc$:
\begin{equation}
f_{\mathrm{int, circles}}\left(T=\Tc,l,r_1,r_2\right)/\left(\kB \Tc\right)=-F_{\mathrm{circle}}\left(r_1\right)F_{\mathrm{circle}}\left(r_2\right)\frac{G\left(0\right)}{l^2}+\mathrm{O}\left(\frac{\ln
  l}{l^3}\right).
\end{equation}
These two results agree, provided that (see Eq.~\eqref{funcG1})
\begin{equation}\label{fcircle}
F_\mathrm{circle}\left(r\right)=-\pi r.
\end{equation}
The sign in this formula is not fixed; we have chosen it such as to comply with our conventions. Since it is not possible to create exactly circular defects on a square lattice, we consider Eq.~\eqref{fcircle} as an approximation valid for large defects.

\begin{figure}
\includegraphics[width=\figwidth]{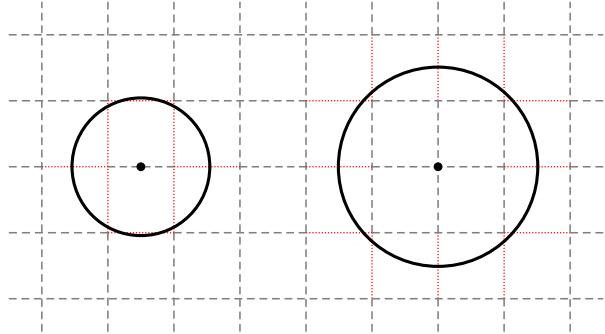}
\caption{\label{fig:circle} Approximation of a circle on a square lattice. Dashed lines correspond to bonds between nearest neighbor spins. This interaction is removed if that bond is crossed by the circumference. These latter bonds are marked in red and are dotted. The two circles shown represent two different types of locations of their centers. The center of the left circle corresponds to the position II (middle of the bond), while the center of the right circle corresponds to the position I (vertex of the lattice).}
\end{figure}

In order to check Eq.~\eqref{fcircle} we calculate the factors in Eq.~\eqref{sl} for defects which mimic circular inclusions which do not interact with the surrounding spins. In order to create such defects we use the following procedure: We choose a position for the center $O$ of a circle and draw a circle of radius $r$. If the circumference crosses a bond connecting a pair of nearest neighbor spins, we delete the interaction between this pair of spins (by setting $J+\Delta J_i=0$). We consider three types of locations of the center of the circle: the center coincides with a vertex of the lattice (position I), it lies in the middle of the bond connecting two neighboring spins (position II), and in the middle of a plaquette, i.e., in the center of a lattice cell (position III). In Fig.~\ref{fig:circle} two examples of this procedure are shown. Upon increasing $r$, the resulting shape of the defect is changing only if the circle intersects a bond connecting a new pair of nearest neighbors. In between these values of $r$, the shape of the defect does not change when $r$ is varied.

\begin{figure}
\includegraphics[width=\figwidth]{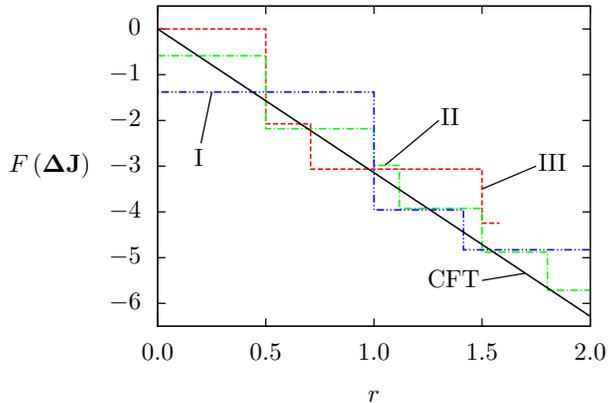}
\caption{\label{fig:circledefectplot}Comparison of the amplitudes $F\left(\mathbf{\Delta J}\right)$ of the interaction free energy $\fint$ (Eq.~\eqref{sl}) for quasi--circular lattice defects with the CFT prediction $F_\mathrm{circle}=-\pi r$ for three positions I--III of the center of the circle as described in the main text.}
\end{figure}

In Fig.~\ref{fig:circledefectplot} we compare the factors in Eq.~\eqref{sl} calculated for the defects with the corresponding value predicted by CFT (see Eq.~\eqref{fcircle}). Due to the numerical complexity of the derivation, we have been able to calculate these values only for $r<2$ for circles with positions I and II and for $r<\sqrt{10}/2$ for circles with the position III. The comparison reveals some degree of similarity but the accessible values of $r$ are not large enough to find good agreement with CFT. However, Fig.~\ref{fig:circledefectplot} gives rise to the expectation that the relative spread $\delta F/F$ vanishes in the limit $r\to\infty$.

\section{Summary and conclusions}\label{sec:sc}

We have considered the two--dimensional Ising model on a square lattice with two lattice defects. Each defect is a group of spins the coupling constants of which differ from the bulk one. For this system, using exact diagonalization of the transfer matrix, we have developed a method of calculating the interaction free energy between these defects. For simple defects (i.e., consisting of a single bond each) we have determined an exact explicit expression for the interaction free energy and the critical Casimir force. For more complicated defects (including the simple ones), we have derived the functional form of the force in the scaling limit (see Eqs.~\eqref{sl} and~\eqref{G}), which we consider as the main result of the present study. The interaction free energy between two defects a distance $l$ apart decays in the scaling limit for large $l$ as $l^{-2}$ and factorizes into three factors. The first factor depends only on the shape and the couplings inside the first defect, the second factor depends only on the shape and the couplings in the second defect, and the third factor is a universal scaling function of the scaling variable $x\sim l/\xib$, where $\xib$ is the bulk correlation length. Finally, we have compared our lattice results with those available in the literature for systems belonging to the same universality class but described in terms of conformal field theory. Our results demonstrate explicitly that in the scaling limit the short--distance expansion (which follows the general idea of the so--called Operator Product Expansion \cite{Cardy87}), applied in Ref.~\cite{Burkhardt95} in order to calculate the interaction free energy of two small spheres at the critical point, extends also to off--critical temperatures.

We note that the leading dependence $\fint\sim l^{-2}$ of the interaction free energy at the critical temperature cannot be determined from a na\"ive dimensional analysis. In the case of the strip geometry at $T=\Tc$, the assumption that the free energy does not depend on microscopic lengthscales together with dimensional analysis renders correctly the free energy per area of the surface to be proportional to $\kB \Tc l^{-\left(d-1\right)}$, where $l$ is the distance between two $\left(d-1\right)$--dimensional parallel plates bounding the $d$--dimensional system. In the present case of two strictly finite--sized defects immersed in a two--dimensional system, the same argument would lead to the incorrect result $\fint\sim \ln l$. Such a logarithmic dependence governs, for example, the interaction free energy between two points pinning the fluctuating interface between two coexisting phases in two dimensions \cite{Abraham07}. Logarithmic dependences of the free energy on the characteristic system size occur also in nonperiodic systems due to the presence of corners \cite{Privman90}.

Due to universality, our results should also apply to proteins immersed in certain lipid bilayers close to their critical demixing point. However, our results cannot be applied directly to actual biological cells because therein the number of proteins is much larger than two and thus many body interactions \cite{Mattos13, Mattos15, Hobrecht15, Paladugu15} must be taken into account. Another problem concerns the size of the system. Our results are valid for macroscopic, flat system while cell membranes are curved and are of finite extent. Therefore, our results can be considered only as a step towards understanding the role of critical Casimir forces in lipid bilayers. 

It would be interesting to study the effective interactions between two defects for the two--dimensional random--field Ising model. In living cells proteins can bind to the underlying cytoskeleton, which leads to a 2D fluid consisting of mobile particles (e.g., lipids) diffusing in a background of quenched (immobilized) protein obstacles. A similar situation arises in supported membranes, for which surface friction (between the lipids and the support structure) may lead to local particle immobilization. As pointed out in Ref.~\cite{Fischer11}, in the presence of quenched obstacles, fluid membranes belong to the universality class of the two--dimensional random--field Ising model. (For critical Casimir forces in the presence of random \textit{surface} fields see Refs.~\cite{ParisenToldin15} and \cite{Maciolek15}.)

\begin{acknowledgments}
The authors thank Erich Eisenriegler for illuminating discussions.
\end{acknowledgments}

\appendix
\section{Transfer matrix for the 2D Ising model}\label{app:A}

All calculations of the free energy have been carried out using the transfer matrix in the horizontal direction $\left(1,0\right)$ (see Fig.~\ref{fig:Sys1}). It has been shown~\cite{Kramers41,Montroll41} that the partition function $\partf_{\text{Ising}}$ of the system without defects for the cyclic boundary conditions imposed in the $(0,1)$ direction can be written in terms of the transfer operator $\opV_2\opV_1$ as
\begin{equation}
 \label{eq:part_func}
 \partf_{\text{Ising}}\left(T,M,N\right) =\sum_{\left\{s_{n,m}\right\}}\exp\left[-\mathcal{H}_{\text{Ising}}/\left(\kB T\right)\right]= (2\sinh 2K)^{NM/2}\Tr\left[\left(\opV_2\opV_1\right)^N\right],
\end{equation}
where $\mathcal{H}_{\text{Ising}}$ is a Hamiltonian of the 2D Ising model without defects (see Eq.~\eqref{HIsing}). The above formula contains the expression
\begin{equation}
\label{eq:v1}
(2\sinh 2K)^{M/2}\opV_1=(2\sinh 2K)^{M/2}\exp\left( -\Kast\sum_{j=0}^{M-1}\sigma _j^z\right),
\end{equation}
with $K=J/\left(\kB T\right)$ and the dual coupling $\Kast$ given by
\begin{equation}\label{Kast}
\sinh 2K \sinh 2\Kast=1,
\end{equation}
which represents the contribution to the Boltzmann factor stemming from the interaction between neighboring rows. At bulk criticality $T=\Tc=J/\left(\kB \Kc\right)$ one has $\Kc=\Kcast=\ln\left(1+\sqrt{2}\right)/2$ so that, for $T\to\Tc$, $K-\Kast\to K_\mathrm{c}-K_\mathrm{c}^\ast=0$. The operator $\opV_2$, which accounts for the Boltzmann factors stemming from interactions within a single column, is determined by horizontal boundary conditions. For cyclic boundary conditions
\begin{equation}
\label{eq:v2}
\opV_2=\exp\left(K\sum_{j=0}^{M-1}\sigma_j^x\sigma_{j+1}^x\right),
\end{equation}
where $\sigma^x_M \equiv \sigma^x_0$. The spin operators $\sigma^{\alpha}_j$, with $\alpha=x, y, z$, operate on a $2^M$ dimensional vector space $X$ which is the tensor product of $M$ two--dimensional vector spaces (each describing the state of a single spin in the column). They are defined by
\begin{equation}
\label{eq:sigmaop}
\sigma^{\alpha}_j=\left(\mathop{\otimes}\limits_{k=0}^{j-1}{\mathbf 1}\right)\otimes\sigma^{\alpha}\otimes\left(\mathop{\otimes}\limits_{k=j+1}^{M-1}{\mathbf 1}\right)
\end{equation}
where $j=0,1,2,\ldots,M-1$, $\sigma^{\alpha}$ are the Pauli spin operators,
\begin{equation}
\sigma^x=\begin{pmatrix} 1 & 0 \\ 0 & -1\end{pmatrix}, \quad
\sigma^y=\begin{pmatrix} 0 & -\iu \\ \iu & 0 \end{pmatrix}, \quad
\sigma^z=\begin{pmatrix} 0 & -1 \\ -1 & 0 \end{pmatrix},
\end{equation}
and ${\mathbf 1}$ is the unit $2\times 2$ matrix. The operators $\sigma_j^\alpha$ fulfill the commutation rule
\begin{equation}
\sigma_j^\alpha \sigma_k^\beta-\sigma_k^\beta \sigma_j^\alpha=\sum_{\gamma}2\iu \delta_{j,k}\varepsilon_{\alpha\beta\gamma}\sigma_j^\gamma,
\end{equation}
where $\delta_{j,k}$ and $\varepsilon_{\alpha\beta\gamma}$ are the Kronecker and the Levi--Civita symbol, respectively.

Using the cyclic property of the trace, Eq.~\eqref{eq:part_func} can be expressed as
\begin{equation}
 \label{eq:part_func_1}
 \partf_{\text{Ising}}\left(T,M,N\right) = (2\sinh 2K)^{NM/2}\Tr\left(\opV^\prime\right)^N
\end{equation}
where $\opV^\prime=\opV_1^{1/2}\opV_2\opV_1^{1/2}$ with $\opV_1^{1/2}=\exp\left(-\left(\Kast/2\right)\sum_{j=0}^{M-1}\sigma _j^z\right)$. Thus, evaluating $\partf_{\text{Ising}}$ is equivalent to finding all eigenvalues of the self--adjoint operator $\opV^\prime$.

Following Ref.~\cite{Kaufman49}, in order to diagonalize $\opV^\prime$ we introduce a set of $2M$ spinors:
\begin{subequations}
\begin{align}
\Gamma_{2j}&=\left[\prod_{k=0}^{j-1}\left(-\sigma^z_k\right)\right]\sigma^x_{j},\\
\Gamma_{2j+1}&=\left[\prod_{k=0}^{j-1}\left(-\sigma^z_k\right)\right]\sigma^y_{j},
\end{align}
\end{subequations}
(with $\Gamma_0=\sigma_0^x$ and $\Gamma_1=\sigma_0^y$) where $j=0,1,2,\ldots,M-1$. These spinors are self--adjoint and satisfy the anti--commutation relation
\begin{equation}
\Gamma_{j}\Gamma_{k}+\Gamma_{k}\Gamma_{j}=2\delta_{j,k}\opI,
\end{equation}
where $\opI$ is the $2^M\times 2^M$ identity matrix. Using spinors, the operators $\opV_1$ and $\opV_2$ can be expressed as
\begin{subequations}
\begin{align}
\label{eq:A:V1}\opV_1&=\exp\left(\rmi \Kast \sum_{j=0}^{M-1}\Gamma_{2j}\Gamma_{2j+1}\right),\\ 
\opV_2&=\exp\left[\rmi K \left(\sum_{j=0}^{M-2}\Gamma_{2j+1}\Gamma_{2j+2}-
  \opPM \Gamma_{2M-1}\Gamma_0\right)\right],
\end{align}
\end{subequations}
where $\opPM=\iu^M \Gamma_0 \Gamma_1 \Gamma_2 \ldots \Gamma_{2M-1}$ is a symmetry operator (i.e., $\opPM^2=\opI$) acting on the vector space $X$. One can verify that $\opPM$ commutes with $\opV_1$ and $\opV_2$, so that these two operators can be diagonalized simultaneously with $\opPM$. We define two Hermitian projection operators $\opP^\pm=\frac{1}{2}\left(\opI\pm\opPM\right)$ and decompose $\opV_2$ as
\begin{equation}
\opV_2=\opP^+ \opV_2^++\opP^-\opV_2^-,
\end{equation}
where 
\begin{equation}\label{V2Gamma}
\opV^\pm_2=\exp\left(\rmi K \sum_{j=0}^{M-2}\Gamma_{2j+1}\Gamma_{2j+2}\mp\rmi K
  \Gamma_{2M-1}\Gamma_0\right).
\end{equation}
The transfer operator is given by
\begin{equation}
\opV^\prime=\opP^+\opV^{\prime +}+\opP^-\opV^{\prime -},
\end{equation}
where 
\begin{equation}\label{Vprimedef}
\opV^{\prime\pm}=\opV_1^{1/2}\opV_2^\pm\opV_1^{1/2}. 
\end{equation}
Since the projection operators $\opP^\pm$ commute also with $\opV^{\prime\pm}$, there exist common eigenvectors of these two operators. The eigenvectors of the transfer operator $\opV^{\prime}$ are the eigenvectors of $\opV^+$ and $\opV^-$ which are unchanged under the action of the projections $\opP^+$ and $\opP^-$, respectively. The problem of finding the spectrum of $\opV^{\prime}$ is now reduced to finding the spectrum of $\opV^{\prime\pm}$ and checking how projection operators $\opP^\pm$ act on their eigenvectors. Since the highest eigenvalue of the transfer operator is an eigenvalue of $\opV^{\prime+}$ \cite{Kaufman49}, below we describe only the procedure of diagonalizing $\opV^{\prime+}$. For $\opV^{\prime-}$ the procedure of diagonalization is very similar.

In order to find the eigenvalues and eigenvectors of $\opV^{\prime+}$ we form a linear transformation of the spinors $\left\{\Gamma_j\right\}$. A new set of anti--commuting spinors $\left\{\opG_j\,|\,j=0,1,2,\ldots, 2M-1\right\}$ is
\begin{subequations}
\begin{equation}
\label{GfromGamma}
\opG_j=\sum_{k=0}^{2M-1}\mS_{kj}\Gamma_k,
\end{equation}
where $\mS$ is a $2M\times 2M$ matrix. In order to have the spinors $\left\{\opG_j\right\}$ being self--adjoint and anti--commuting, the matrix $\mS$ must be real and orthogonal and thus
\begin{equation}\label{GammafromG}
 \Gamma_j=\sum_{k=0}^{2M-1}\mS_{jk}\opG_k.
\end{equation}
\end{subequations}
Following Refs.~\cite{Kaufman49, Huang63} we take
\begin{subequations}
\label{eq:A:S}
\begin{align}
\mS_{j,2k-1}&=\sqrt{2}\operatorname{Re}y_j^{\left(k\right)}, & \mS_{j,2k}&=\sqrt{2}\operatorname{Im}y_j^{\left(k\right)},\\
\mS_{j,0}&=\sqrt{2}\operatorname{Im}y_j^{\left(0\right)}, &
\mS_{j,2M-1}&=\sqrt{2}\operatorname{Re}y_j^{\left(0\right)},
\end{align}
where
\begin{align}
y^{\left(k\right)}_{2l}&=N_k\ee^{\iu\omega_k
  l}\left(\cosh\Kast+\iu\,q_k \sinh\Kast\right), &y^{\left(k\right)}_{2l+1}&=N_k\ee^{\iu\omega_k
  l}\left(-\iu \sinh\Kast+q_k\cosh\Kast\right),\\
N_k&=\left(\frac{\ee^{\gamma_k}\cosh 2\Kast-\cosh
    2K}{2M\sinh\gamma_k}\right)^{1/2}, & q_k&=\iu \frac{\ee^{\gamma_k}\cosh
  2\Kast-\cosh 2K}{\ee^{\gamma_k}\sinh 2\Kast-\ee^{-\iu \omega_k}\sinh 2K},
\end{align}
$j=0,1,2\ldots,2M-1$; $k,l=0,1,2,\ldots, M-1$; and
\begin{equation}
\label{eq:A:gamma}
\gamma_k=\gamma\left(\omega_k\right)=\operatorname{arccosh}\left(\cosh 2\Kast\cosh 2K-\cos\omega_k\right),\quad \gamma_k>0,\quad \omega_k=\left(2k+1\right)\pi/M.
\end{equation}
\end{subequations}
The matrix $\mS$ was chosen such that the transfer operator $\opV^{\prime+}$ expressed in terms of the set $\left\{\opG_j\right\}$ of spinors has the simple form
\begin{equation}
\label{eq:A:VwithG}
\opV^{\prime+}=\exp\left(\frac{\iu}{2}\sum_{k=1}^{M-1}\gamma_k\opG_{2k-1}\opG_{2k}+\frac{\iu}{2}\gamma_0\opG_{2M-1}\opG_0\right).
\end{equation}
In the next step we introduce the fermionic annihilation operators
\begin{equation}
\label{eq:A:Gtof}
f_0=\frac{1}{2}\left(\opG_0+\iu\opG_{2M-1}\right), \quad
f_k=\frac{1}{2}\left(\opG_{2k}+\iu\opG_{2k-1}\right),\quad k=1,2,\ldots,M-1.
\end{equation}
Using these operators we obtain
\begin{equation}
\opV^{\prime+}=\exp\left[-\sum_{k=0}^{M-1}\gamma_k\left(f^\dagger_kf_k-\frac{1}{2}\opI\right)\right].
\end{equation}
This formula shows that the occupation number basis defined by the fermionic operators is an eigenbasis of $\opV^{\prime+}$. Because all coefficients $\gamma_k$ are positive, the eigenvector $\left|0\right>$ corresponding to the largest eigenvalue $\Lambda_0$ satisfies
\begin{equation}
\label{eq:A:vacuum}
f_k\left|0\right>=0,\qquad \text{for all}\quad k=0,1,2,\ldots,M-1,
\end{equation}
and
\begin{equation}
\Lambda_0=\exp\left(\sum_{k=0}^{M-1}\gamma_k/2\right).
\end{equation}
All the other eigenvectors and eigenvalues follow from
\begin{equation}
\left|L\right>=f_{l_n}^\dagger f_{l_{n-1}}^\dagger\ldots
f_{l_1}^\dagger\left|0\right>, \qquad \opV^{\prime+}\left|L\right>=\Lambda_0\exp\left(-\gamma_{l_1}-\gamma_{l_2}-\ldots-\gamma_{l_n}\right)\left|L\right>,
\end{equation}
where $L=\left\{l_1,l_2,\ldots,l_n\right\}$ and $0\leqslant l_1<l_2<\ldots<l_n<M$. Additionally, the projection operator $\opP^+$ acting on the eigenvectors yields \cite{Kaufman49}
\begin{equation}
\opP^+\left|L\right>=\left|L\right> \quad \text{for $n$ even},\qquad
\opP^+\left|L\right>=0 \quad \text{for $n$ odd}.
\end{equation} 
In particular, this implies $\opP^+\left|0\right>=\left|0\right>$ and this vector, associated with the highest eigenvalue, is an eigenvector of the full transfer operator $\opV^\prime$.

These canonical formulae must be modified in order to incorporate the lattice defects. This is accomplished by introducing a special operator $\opD$ which changes the couplings between selected pairs of spins. The method for constructing such an operator is described in Appendix \ref{app:B}. The operator $\opD$ is defined via
\begin{equation}\label{def_partf}
\partf=\left(2 \sinh 2K\right)^{NM/2}\operatorname{Tr}\left[\opD \left(\opV^{\prime}\right)^N\right],
\end{equation}
where $\partf$ is the partition function of the system with defects and $\opV^{\prime}$ is the transfer operator for the system without defects (see Eq.~\eqref{eq:part_func_1}). In the thermodynamic limit $M,N\to\infty$ it is sufficient to consider only the highest eigenvalue of the transfer operator $\opV^\prime$ (see Appendix~\ref{app:C}) so that
\begin{equation}
\partf\approx\left(2 \sinh 2K\right)^{NM/2} \left(\Lambda_0\right)^N\left<0\middle| \opD\middle| 0\right>.
\end{equation}

\section{Modification of couplings}\label{app:B}

The operator $\opD$ which modifies couplings between spins inside a defect is constructed as a product of operators $\opR_k$ and $\opC_k$ which modify a single coupling between neighboring spins within one row or one column.

The operator $\opR_k\left(\Delta J\right)$ is changing the coupling between neighboring spins in the $k$-th row. This operator is sandwiched between two transfer operators $\opV^{\prime+}$ which deal with the interaction of spins in two adjacent columns. This means that $\opR_k$ must satisfy
\begin{equation}\label{defR}
\opV^{\prime+}\ \opR_k\left(\Delta J\right)\opV^{\prime+}=\opV_1^{1/2}\ \opV_2^+
\ \tilde{\opV}_1\left(\Delta J\right)\opV_2^+\ \opV_1^{1/2},
\end{equation}
where the operator $\tilde{\opV}_1$ is the operator $\opV_1$ given by Eq.~\eqref{eq:v1} with a modified coupling $J+\Delta J$ between the spins in the $k$-th row:
\begin{multline}
(2\sinh 2K)^{M/2}\tilde{\opV}_1\left(\Delta J\right)=\\
(2\sinh 2K)^{\left(M-1\right)/2}\left[2\sinh\left(2K+2\Delta
  K\right)\right]^{1/2}\exp\left[ -\Kast\sum_{j=0,j\neq
    k}^{M-1}\sigma_j^z-\left(K+\Delta K\right)^\ast\sigma_k^z\right],
\end{multline}
where $\Delta K=\Delta J/\left(\kB T\right)$ and $\left(K+\Delta K\right)^\ast$ is defined as discussed in Appendix~\ref{app:A}.

Simplifying Eq.~\eqref{defR} leads to
\begin{equation}\label{opRdef}
\opR_k\left(\Delta J\right)=\opV_1^{-1/2}\tilde{\opV}_1\opV_1^{-1/2} =\opI\ \frac{\sinh\left(2K+\Delta K\right)}{\sinh 2K}-\rmi\,
\Gamma_{2k}\Gamma_{2k+1}\frac{\sinh\Delta K}{\sinh 2K}.
\end{equation}
This calculation requires special care in the case $K+\Delta K<0$, in which both $\left(K+\Delta K\right)^\ast$ and $\left[2\sinh\left(2K+2\Delta K\right)\right]^{1/2}$ are complex.

The operator $\opC_k$ is changing the coupling of neighboring spins in one column, located in $k$-th and $\left(k+1\right)$-th row, respectively. We use the convention that this operator is put to the right of the transfer matrix $\opV^{\prime+}$ of the column to be modified:
\begin{equation}\label{cdef}
\opV^{\prime+}\opC_k\left(\Delta
  J\right)=\opV_1^{1/2}\ \tilde{\opV}_2^+\left(\Delta J\right)\opV_1^{1/2},
\end{equation}
where $\tilde{\opV}_2^+\left(\Delta J\right)$ is an operator $\opV_2$ as given by Eq.~\eqref{eq:v2} but with a modified coupling $J+\Delta J$ between the spins in the $k$-th and in the $\left(k+1\right)$-th row:
\begin{equation}
\tilde{\opV}_2^+\left(\Delta J\right)=\opP^+\exp\left[K\sum_{j=0,j\neq k}^{M-1}\sigma_j^x\sigma_{j+1}^x+\left(K+\Delta K\right)\sigma_k^x\sigma_{k+1}^x\right],
\end{equation}
which for $k\neq M-1$ is
\begin{equation}\label{V2tilde}
\tilde{\opV}_2^+\left(\Delta J\right)=\exp\left(\rmi
  K\sum_{j=0}^{M-2}\Gamma_{2j+1}\Gamma_{2j+2}-\rmi K
  \Gamma_{2M-1}\Gamma_0+\rmi \Delta K \Gamma_{2k+1}\Gamma_{2k+2}\right).
\end{equation}
From the definition of $\opV^{\prime+}$ in Eq.~\eqref{Vprimedef} and due to $\left[\opV_1^{1/2}\opV_2^+ \opV_1^{1/2}\right]^{-1}=\opV_1^{-1/2}\left(\opV_2^+\right)^{-1}\opV_1^{-1/2}$ we obtain from Eq.~\eqref{cdef}
\begin{equation}
\opC_k\left(\Delta
  J\right)=\opV_1^{-1/2}\left(\opV_2^+\right)^{-1}\tilde{\opV}_2^+\opV_1^{1/2}.
\end{equation}
Using the formulae in Eqs.~\eqref{eq:A:V1}, \eqref{V2Gamma}, and \eqref{V2tilde}, and certain commutations of the $\Gamma_k$ spinors leads to
\begin{multline}\label{opCdef}
\opC_k\left(\Delta J\right)=\opI\cosh\Delta K+\rmi \sinh\Delta
K\Big(\Gamma_{2k+1}\Gamma_{2k+2}\cosh^2\Kast
+\Gamma_{2k}\Gamma_{2k+3}\sinh^2
\Kast\\+\rmi\,\Gamma_{2k+1}\Gamma_{2k+3}\sinh\Kast\cosh\Kast-\rmi\,\Gamma_{2k}\Gamma_{2k+2}\sinh\Kast\cosh\Kast\Big).
\end{multline}
This formula does not hold for $k=M-1$. Since, however, any defect of finite size can be located on the lattice such, that the coupling between spins in the $(M-1)$-th row is unmodified, here we do not discuss the expression for the operator $\opC_{M-1}$.

If more than one bond is modified, two or more operators $\opR_k$ and $\opC_k$ must be used. Although for any $k,l=0,1,\ldots, M-1$ the operators commute among themselves, i.e., $\left[\opR_k,\opR_l\right]=\left[\opC_k,\opC_l\right]=0$, the mutual commutators $\left[\opR_k,\opC_l\right]$ can be nonzero. Therefore care must be taken for placing these operators in the correct order. From Eqs.~\eqref{opRdef} and \eqref{opCdef} it follows that only $\opC_k$ depends on the form of the row -- row interaction operator $\opV_1$, which is modified by $\opR_k$. Therefore, all operators $\opC_k$ must act prior to $\opR_k$.

If the defect spans across more than one column, within the corresponding matrix products the transfer operators $\opV^{\prime+}$ are positioned between the operators $\opC$ and $\opR$.

In order to compactify the expression for the operator generating an entire defect, we introduce the following notation:
\begin{equation}
\label{eq:B:ndef}
\shift{n}{\opA}=\left(\opV^{\prime+}\right)^{n}\opA\left(\opV^{\prime+}\right)^{-n},
\end{equation}
where $\opA$ is an arbitrary operator acting on $X$. This way, all operators modifying interactions in the $n$-th column can be shifted to the left side of the product of transfer operators in a matrix element, carrying an upper index $\left(n\right)$, and the formula for the partition function $\partf$ of the system with defects can be expressed in the form stated in Eq.~\eqref{def_partf}. We note that $\shift{n}{\Gamma}_k$ is a linear combination of spinors $\left\{\Gamma_l\right\}$.

The commutation rules for operators $\opC_k$ and $\opR_k$ imply the following ordering: Within the corresponding product of operators, the indices of columns must increase from left to right and within a group of operators with the same upper index, all operators $\opC_k$ act prior to $\opR_k$.

\section{Interaction free energy}\label{app:C}

In this appendix we derive the expression in Eq.~\eqref{fint} for the interaction free energy between two defects. 

First, we comment on the thermodynamic limit. We denote the highest eigenvalue of $\opV^{\prime-}$ by $\tilde{\Lambda}_0$ and its eigenvector by $\left|\tilde{0}\right>$. For $T<\Tc$ this eigenvector is also an eigenvector of $\opV^\prime$ and, due to $\lim_{M\to\infty}\Lambda_0=\lim_{M\to\infty}\tilde{\Lambda}_0$, the spectrum of $\opV^\prime$ is asymptotically degenerated \cite{Kaufman49, Schultz64}. This leads to an extra $M$-- and $N$--independent term $-\kB T \ln 2$ contributing to the free energy of the system. However, due to $\lim_{M\to\infty}\left<0\middle|\opD\middle|0\right>=\lim_{M\to\infty}\left<\tilde{0}\middle|\opD\middle|\tilde{0}\right>$, for any operator $\opD$ which modifies couplings (with the exception of $\opC_{M-1}$ which is not considered here), this degeneracy does neither affect the free energies of single defects nor the interaction free energy. In order to simplify the calculations, instead of the standard thermodynamic limit $M,N\to\infty$ with $0<M/N<\infty$ fixed, we take the limit $N\to\infty$ followed by the limit $M\to\infty$. Since for any finite $M$ one has $\Lambda_0>\tilde{\Lambda}_0$, this approach avoids to consider eigenvectors of $\opV^{\prime-}$; however, the free energy is changed (relative to the one obtained via the standard approach) by an $M$-- and $N$--independent term.

The free energy of the system without (0) defects can be decomposed as follows:
\begin{multline}\label{decomposition1}
\fe_0\left(T,M,N\right)=-\frac{1}{2}\kB T N M \ln \left(2\sinh2K\right)-\kB T \ln
\operatorname{Tr}\left[\left(\opV^{\prime}\right)^N\right]=\\
 N M f_\mathrm{b}\left(T\right)+f_{\text{finite},0}\left(T,M,N\right)+f_0\left(T\right),
\end{multline}
where $f_\mathrm{b}$ is the bulk free energy density (i.e., per spin) of the 2D Ising model, $f_{\text{finite},0}$ is the finite size correction which vanishes in the thermodynamic limit and $f_0$ is the aforementioned $M$-- and $N$--independent term (which depends on how the thermodynamic limit is taken; within our approach $f_0\left(T\right)=0$). Equation~\eqref{decomposition1} defines $f_\mathrm{b}$, $f_{\text{finite},0}$, and $f_0$. Since in the limit $N\to\infty$, $\operatorname{Tr}\left[\left(\opV^{\prime}\right)^N\right]\approx\left(\Lambda_0\right)^N$, our expression for the free energy per spin,
\begin{equation}
f_\mathrm{b}\left(T\right)=-\kB T\left[\frac{1}{2} \ln\left( 2\sinh 2K\right)+ \lim_{M\to\infty}\frac{1}{M}\ln\Lambda_0\right],
\end{equation}
agrees with known results \cite{Onsager44}.

If there is only a single (1) defect in the system, its free energy can be decomposed as
\begin{multline}\label{fIi}
\fe_{1,i}\left(T,\mathbf{\Delta J}_i,M,N\right)=-\frac{1}{2}\kB T N M \ln \left(2\sinh 2K\right)-\kB T\ln \operatorname{Tr}\left[\opD_i
  \left(\opV^{\prime}\right)^N\right]=\\
M N f_\mathrm{b}\left(T\right)+f_i\left(T,\mathbf{\Delta J}_i\right)+f_{\text{finite},1,i}\left(T,\mathbf{\Delta J}_i,M,N\right)+f_0\left(T\right),
\end{multline}
where $i=\alpha,\beta$ labels the defect, $\opD_\alpha\left(\mathbf{\Delta J}_\alpha\right)$ and $\opD_\beta\left(\mathbf{\Delta J}_\beta,l\right)$ are operators which generate the defects by modifying couplings, $f_i$ is the free energy of the defect (in excess of $MNf_\mathrm{b}$), and $f_{\text{finite},1,i}$ is the finite size contribution. If $N$ is large, one has $\operatorname{Tr}\left[\opD_i \left(\opV^{\prime}\right)^N\right]\approx \left(\Lambda_0\right)^N\left<0\middle|\opD_i\middle|0\right>$. Thus the comparison of Eqs.~\eqref{decomposition1} and \eqref{fIi} leads to the conclusion that the free energy of a single defect $i$ is given by
\begin{equation}\label{fi}
f_i\left(T,\mathbf{\Delta
    J}_i\right)=\lim_{M\to\infty}\lim_{N\to\infty}\left[\fe_{1,i}\left(T,M,N,\mathbf{\Delta
      J}_i\right)-\fe_0\left(T,M,N\right)\right]=-\kB T \lim_{M\to\infty}\ln \left<0\middle|\opD_i\middle|0\right>.
\end{equation}
We note that although the operator $\opD_\beta$ depends on $l$, the free energy $f_\beta$ of the second defect $\beta$ is independent of $l$.

The free energy of the system with two (2) defects ($\alpha$ and $\beta$) can be decomposed as
\begin{multline}\label{f2}
\fe_{2}\left(T,l,\mathbf{\Delta J}_\alpha,\mathbf{\Delta J}_\beta,M,N\right)=-\frac{1}{2}\kB T N M \ln \left(2\sinh 2K\right)-\kB T
\ln\operatorname{Tr}\left[\opD_\alpha\opD_\beta\left(\opV^{\prime}\right)^N\right]=\\
NM f_\mathrm{b}\left(T\right)+f_\alpha\left(T,\mathbf{\Delta
    J}_\alpha\right)+f_\beta\left(T,\mathbf{\Delta
    J}_\beta\right)+\fint\left(T,l,\mathbf{\Delta J}_\alpha,\mathbf{\Delta
    J}_\beta\right)\\
+f_{\text{finite},2}\left(T,l,\mathbf{\Delta
    J}_\alpha,\mathbf{\Delta J}_\beta,M,N\right)+f_0\left(T\right),
\end{multline}
where $\fint$ is the interaction free energy and $f_{\text{finite},2}$ is the finite--size contribution. If $N$ is large, one has $\operatorname{Tr}\left[\opD_\alpha\opD_\beta\left(\opV^{\prime}\right)^N\right]\approx \left(\Lambda_0\right)^N\left<0\middle|\opD_\alpha \opD_\beta\middle|0\right>$. From Eqs.~\eqref{f2}, \eqref{decomposition1}, and \eqref{fi} one obtains the following expression for the interaction free energy:
\begin{multline}
\fint\left(T,l,\mathbf{\Delta J}_\alpha,\mathbf{\Delta
    J}_\beta\right)=\\
\lim_{M\to\infty}\lim_{N\to\infty}\left[\fe_{2}\left(T,l,\mathbf{\Delta
      J}_\alpha,\mathbf{\Delta
      J}_\beta,M,N\right)-\fe_0\left(T,M,N\right)-f_\alpha\left(T,\mathbf{\Delta
    J}_\alpha\right)-f_\beta\left(T,\mathbf{\Delta J}_\beta\right)\right]=\\
-\kB T \ln\lim_{M\to\infty}\left<
  0\middle| \opD_\alpha\opD_\beta\middle|
  0\right>+\kB T \lim_{M\to\infty}\ln \left<0\middle|\opD_\alpha\middle|0\right>+\kB T \lim_{M\to\infty}\ln \left<0\middle|\opD_\beta\middle|0\right>
\end{multline}
which leads directly to Eq.~\eqref{fint}.

\section{Calculation of matrix elements}\label{app:D}

In the previous sections and appendices we have discussed how to construct operators $\opD$ which describe lattice defects. The corresponding partition function requires to calculate matrix element $\left<0\middle| \opD\middle| 0\right>$. The operator $\opD$ is a multinomial of spinors of the form $\shift{n}{\Gamma}_j$. Thus by using linearity and Wick's theorem \cite{Wick50}, the matrix element can be decomposed as a sum of Pfaffians of matrix elements of the form $\left<0\middle|\shift{n_1}{\Gamma}_j\shift{n_2}{\Gamma}_k\middle|0\right>$. In order to derive them, we first note that
\begin{equation}\label{two_shifts}
\left<0\middle|
  \shift{n_1}{\Gamma}_j\shift{n_2}{\Gamma}_k\middle|
  0\right>=\left<0\middle|\Gamma_j\shift{n_2-n_1}{\Gamma}_k\middle|
  0\right>,
\end{equation} 
and define
\begin{equation}\label{Wdef}
\mew{j}{k}{n}=\lim_{M\to\infty}\iu\left<0\middle|
  \Gamma_j\shift{n}{\Gamma}_k\middle|
  0\right>,
\end{equation}
where $j,k=0,1,2,\ldots, 2M-1$ and $n$ is an integer. These matrix elements are calculated using the formulae derived in Appendix~\ref{app:A} for the defect free 2D Ising model. With Eqs.~\eqref{eq:A:VwithG} and \eqref{eq:B:ndef} we obtain
\begin{align}
\shift{n}{\opG}_{2k-1}&=\cosh \left(n\gamma_k\right) \,\opG_{2k-1}-\iu \sinh \left(n\gamma_k\right)\,\opG_{2k},\\
\shift{n}{\opG}_{2k}&=\cosh \left(n\gamma_k\right) \,\opG_{2k}+\iu \sinh \left(n\gamma_k\right)\,\opG_{2k-1},
\end{align}
with $\gamma_k$ given by Eq.~\eqref{eq:A:gamma}. The above formulae, together with Eqs.~\eqref{GfromGamma} and \eqref{eq:A:Gtof}, yield
\begin{equation}
\mew{j}{k}{n}=\sum_{m=0}^{M-1}\left(\iu\,\mS_{j,2m}\mS_{k,2m}+\iu\,\mS_{j,2m-1}\mS_{k,2m-1}-\mS_{j,2m}\mS_{k,2m-1}+\mS_{j,2m-1}\mS_{k,2m}\right)\ee^{-n\gamma_m},
\end{equation}
where we assume $\mS_{k,-1}\equiv \mS_{k,2M-1}$. Using Eq.~\eqref{eq:A:S} one can check the relations
\begin{subequations}\label{meW}
\begin{gather}
\mew{2j}{2k}{n}=\mew{2j+1}{2k+1}{n}=\mew{2k}{2j}{n}=\mew{0}{2k-2j}{n},\\
\mew{2j}{2k+1}{n}=-\mew{2j+1}{2k}{n}=\mew{0}{2k+1-2j}{n},
\end{gather}
which tell, inter alia, that every matrix element $\mew{j}{k}{n}$ can be transformed to the one with $j=0$. By carrying out the limit $M\to\infty$ one finds
\begin{gather}
\mew{0}{2k}{n}=\frac{\iu}{\pi}\int_{0}^{\pi}\exp\left[-n
  \gamma\left(\omega,T\right)\right]\cos \left(k\omega\right)\, \dd \omega,\\
\begin{align}
\nonumber\mew{0}{2k+1}{n}=\frac{1}{\pi}\int_0^\pi\exp\left[-n\gamma\left(\omega\right)\right][\left(\ctgh 2K-\cosh 2K
  \cos\omega\right)\cos \left(k\omega\right)\\
+\sinh 2K\sin\omega \sin \left(k\omega\right)]/\sinh\gamma\left(\omega\right)\dd \omega,
\end{align}
\end{gather}
\end{subequations}
where $\gamma\left(\omega\right)$ is given by Eq.~\eqref{eq:A:gamma}.

\section{Critical Casimir force in the scaling limit}\label{app:E}

In this appendix we prove that the interaction free energy between two defects of arbitrary shape exhibits the form given by Eq.~\eqref{scalesimple}. For simplicity, by ``$\text{SL}$'' we denote the scaling limit $T\to\Tc$ and $l\to\infty$ with $x=t l/\xi_0^+$ fixed (see Eq.~\eqref{scalex}). We use the symbol ``$\bullet$'' in order to indicate that after calculating the matrix element, the limit $M\to\infty$ is carried out. We consider only defects which do not change size, strength, and shape in the scaling limit.

As a first step of the proof, we note that
\begin{equation}\label{Wscale}
\lim_{\text{SL}}l\,
\mew{0}{2k+1}{l+a}=\frac{1}{\pi}x\BesselK_0\left(\left|x\right|\right),
\qquad \lim_{\text{SL}}l\,\mew{0}{2k}{l+a}=\frac{\iu}{\pi} x \BesselK_1\left(\left|x\right|\right).
\end{equation}
The limits are independent of $k$ and $a$, and $\BesselK_j$ are modified Bessel functions of the second kind. Equation \eqref{Wscale} follows directly from Eq.~\eqref{meW} in the scaling limit.

In the second step, we consider two operators $\opD_{\alpha}$ and $\opD_{\beta}$ which represent the defects. They consist of a product of operators $\shift{n}{\opC}_k$ and $\shift{n}{\opR}_k$ and thus can be expressed in the general form
\begin{equation}
\opD_{\alpha}=f_0\opI+\sum_{a}f_a \opA_a, \qquad \opD_{\beta}=g_0\opI+\sum_{b}g_b \opB_b,
\end{equation}
where $f_a$ and $g_b$ are scalars which depend on temperature and the coupling constants in each defect. $\opA_a$ and $\opB_b$ are products of \textit{even} numbers of spinors of the form $\shift{n}{\Gamma}_i$ and $\shift{l+n}{\Gamma}_i$, respectively.

The linearity of matrix elements implies
\begin{equation}\label{D1D2}
\left<0\middle| \opD_{\alpha} \opD_{\beta}\middle| 0\right>^\bullet=f_0 g_0+\sum_a g_0 f_a
\left<0\middle|\opA_a\middle|0\right>^\bullet+\sum_b f_0 g_b\left<0\middle|
  \opB_b\middle|0\right>^\bullet+\sum_a\sum_bf_a g_b
\left<0\middle| \opA_a \opB_b\middle| 0\right>^\bullet.
\end{equation}
We now consider one particular matrix element contributing to the last sums in Eq.~\eqref{D1D2}. The general form of $\opA_a$ and $\opB_b$ is
\begin{equation}
\opA_a=\rmi^P\shift{m_1}{\Gamma}_{p_1}\shift{m_2}{\Gamma}_{p_2}\ldots\shift{m_{2P}}{\Gamma}_{p_{2P}},\qquad
\opB_b=\rmi^Q\shift{l+n_1}{\Gamma}_{q_1}\,\shift{l+n_2}{\Gamma}_{q_2}\ldots\shift{l+n_{2Q}}{\Gamma}_{q_{2Q}},
\end{equation}
where $\left\{m_j\right\}$ and $\left\{p_j\right\}$ are sets of $2P$ natural numbers, and $\left\{n_j\right\}$ and $\left\{q_i\right\}$ are sets of $2Q$ natural numbers. According to Wick's theorem, in order to calculate $\left<0\middle| \opA_a \opB_b\middle| 0\right>^\bullet$ one must consider all possible pair contractions of the $\Gamma$--spinors in the operator $\opA_a \opB_b$. If there are no contractions connecting spinors belonging to $\opA_a$ and $\opB_b$ the resulting term does not depend on $l$. All terms of this latter kind sum up to $\left<0\middle| \opA_a\middle| 0\right>^\bullet\left<0\middle| \opB_b\middle| 0\right>^\bullet$. Since there is an even number of spinors in $\opA_a$ and $\opB_b$, there is no term with only one contraction between spinors from $\opA_a$ and $\opB_b$. When there are more than two contractions between spinors from $\opA_a$ and $\opB_b$, according to Eq.~\eqref{Wscale} the resulting term decays faster than $l^{-2}$ in the scaling limit ($\lim_{\text{SL}}$). Thus the leading order of the dependence on $l$ is generated by terms with precisely two contractions between spinors from two distinct operators. The sum of all of these terms is given by
\begin{multline}\label{SLformula}
  \lim_{\text{SL}}l^2\Big[\left<0\middle|\opA_a\opB_b\middle|0\right>^\bullet -
    \left<0\middle|\opA_a\middle|0\right>^\bullet
    \left<0\middle|\opB_b\middle|0\right>^\bullet\Big] =\\
-\lim_{\text{SL}}
  \underset{\mu<\nu}{\sum_{\mu,\nu=1}^{2P}} \,
  \underset{\rho<\sigma}{\sum_{\rho,\sigma=1}^{2Q}}
  \left(-1\right)^{\mu+\nu+\rho+\sigma}l^2\left(\mew{p_\mu}{q_\rho}{l+n_\rho-m_\mu}\mew{p_\nu}{q_\sigma}{l+n_\sigma-m_\nu}-
  \mew{p_\mu}{q_\sigma}{l+n_\sigma-m_\mu}\mew{p_\nu}{q_\rho}{l+n_\rho-m_\nu}\right)\\
  \times\left<0\middle|\rmi^{P-1}\shift{m_1}{\Gamma}_{p_1}
    \shift{m_2}{\Gamma}_{p_2}\ldots
    \shift{m_{\mu-1}}{\Gamma}_{p_{\mu-1}}
    \shift{m_{\mu+1}}{\Gamma}_{p_{\mu+1}}
    \ldots\shift{m_{\nu-1}}{\Gamma}_{p_{\nu-1}}
    \shift{m_{\nu+1}}{\Gamma}_{p_{\nu+1}} \ldots\shift{m_{2P}}{\Gamma}_{p_{2P}}
    \middle|0\right>^\bullet\\
\times\left<0\middle|\rmi^{Q-1}\shift{l+n_1}{\Gamma}_{q_1}\,\shift{l+n_2}{\Gamma}_{q_2}
  \ldots\shift{l+n_{\rho-1}}{\Gamma}_{q_{\rho-1}}\,\shift{l+n_{\rho+1}}{\Gamma}_{q_{\rho+1}}
  \ldots\shift{l+n_{\sigma-1}}{\Gamma}_{q_{\sigma-1}}\,\shift{l+n_{\sigma+1}}{\Gamma}_{q_{\sigma+1}}
  \ldots\shift{l+n_{2Q}}{\Gamma}_{q_{2Q}}
    \middle|0\right>^\bullet,
\end{multline}
where both possible contractions between $\shift{m_\mu}{\Gamma}_{p_\mu}\shift{m_\nu}{\Gamma}_{p_\nu}$ and $\shift{l+n_\rho}{\Gamma}_{q_\rho}\shift{l+n_\sigma}{\Gamma}_{q_\sigma}$ have been expressed in terms of matrix elements $\meW$ using Eq.~\eqref{Wdef}. Since the contractions between two spinors of the form $\shift{l+a}{\Gamma}_{b}$ do not depend on $l$ (see Eq.~\eqref{two_shifts}), the matrix element, which is the last factor in Eq.~\eqref{SLformula}, is $l$--independent. In order to further simplify Eq.~\eqref{SLformula} we note that
\begin{multline}
\lim_{\text{SL}}l^2\left(\mew{p_\mu}{q_\rho}{l+n_\rho-m_\mu}
  \mew{p_\nu}{q_\sigma}{l+n_\sigma-m_\nu}-
  \mew{p_\mu}{q_\sigma}{l+n_\sigma-m_\mu} \mew{p_\nu}{q_\rho}{l+n_\rho-m_\nu}\right)=\\-\frac{1}{4}
\left[\left(-1\right)^{p_\mu}-\left(-1\right)^{p_\nu}\right]
\left[\left(-1\right)^{q_\rho}-\left(-1\right)^{q_\sigma}\right]
G\left(x\right),
\end{multline}
where the prefactor (in square brackets) of $G\left(x\right)=x^2\left[\BesselK_1^2\left(\left|x\right|\right)-\BesselK_0^2\left(\left|x\right|\right)\right]/\pi^2$ (see Eq.~\eqref{Wscale}) is non--zero only if $p_\mu+p_\nu$ as well as $q_\rho+q_\sigma$ are both odd in which case it is equal to $+1$ or $-1$. Note that the above limit does not depend on $n_\rho$, $n_\sigma$, $m_\mu$ and $m_{\nu}$; this follows from Eq.~\eqref{Wscale}.

Since the remaining factors in Eq.~\eqref{SLformula} are independent of $l$, the scaling limit can be replaced by the limit $T\to\Tc$. Simple transformations yield
\begin{multline}
  \lim_{\text{SL}}l^2\Big[\left<0\middle|\opA_a\opB_b\middle|0\right>^\bullet -
    \left<0\middle|\opA_a\middle|0\right>^\bullet
    \left<0\middle|\opB_b\middle|0\right>^\bullet\Big]=\\
G\left(x\right)\underset{\text{$p_\mu$ even, $q_\nu$
    odd}}{\sum_{\mu,\nu=1}^{2P}}
\left(-1\right)^{\mu+\nu}\operatorname{sign}
\left(\nu-\mu\right)\\
\times\left. \left<0\middle|\rmi^{P-1}\shift{m_1}{\Gamma}_{p_1}
    \shift{m_2}{\Gamma}_{p_2}\ldots\shift{m_{\mu-1}}{\Gamma}_{p_{\mu-1}}
    \shift{m_{\mu+1}}{\Gamma}_{p_{\mu+1}}\ldots
    \shift{m_{\nu-1}}{\Gamma}_{p_{\nu-1}}
    \shift{m_{\nu+1}}{\Gamma}_{p_{\nu+1}} \ldots \shift{m_{2P}}{\Gamma}_{p_{2P}}
    \middle|0\right>^\bullet\right|_{T=\Tc}\\
\times\underset{\text{$q_\rho$ even, $q_\sigma$
      odd}}{\sum_{\rho,\sigma=1}^{2Q}}\left(-1\right)^{\rho+\sigma} \operatorname{sign}\left(\sigma-\rho\right)\\
\times\left.\left<0\middle|\rmi^{Q-1}\shift{n_1}{\Gamma}_{q_1}\,\shift{n_2}{\Gamma}_{q_2}
      \ldots\shift{n_{\rho-1}}{\Gamma}_{q_{\rho-1}}\,\shift{n_{\rho+1}}{\Gamma}_{q_{\rho+1}}
      \ldots\shift{n_{\sigma-1}}{\Gamma}_{q_{\sigma-1}}\,
      \shift{n_{\sigma+1}}{\Gamma}_{q_{\sigma+1}}\ldots\shift{n_{2Q}}{\Gamma}_{q_{2Q}}
    \middle|0\right>^\bullet\right|_{T=\Tc}=\\
G\left(x\right)\left<0\middle|\tilde{\opA}_a\middle|0\right>
    \left<0\middle|\tilde{\opB}_b\middle|0\right>,
\end{multline}
where we have introduced the quantities $\left<0\middle|\tilde{\opA}_a\middle|0\right>$ and $\left<0\middle|\tilde{\opB}_b\middle|0\right>$ for the above sums of matrix elements. Using this result in Eq.~\eqref{D1D2} gives
\begin{multline}
\lim_{\text{SL}}l^2\Big[\left<0\middle|\opD_{\alpha}\opD_{\beta}\middle|0\right>^\bullet-
  \left<0\middle|\opD_{\alpha}\middle|0\right>^\bullet \left<0\middle|
    \opD_{\beta}\middle|0\right>^\bullet\Big]=\\
\sum_a\sum_b f_a g_b
\lim_{\text{SL}}l^2\big[ \left<0\middle|\opA_a\opB_b\middle|0\right>^\bullet-
  \left<0\middle|\opA_a\middle|0\right>^\bullet \left<0\middle|
    \opB_b\middle|0\right>^\bullet\big]=\\
G\left(x\right)\left(\sum_a f_a
  \left<0\middle|\tilde{\opA}_a\middle|0\right>\right) \left(\sum_b g_b
  \left<0\middle|\tilde{\opB}_b\middle|0\right> \right).
\end{multline}
Inserting this expression into Eq.~\eqref{fint} and omitting higher order terms in the scaling limit leads to the factorization stated in Eq.~\eqref{sl}.


%
\end{document}